\documentclass{emulateapj}
\usepackage{graphicx}
\usepackage{natbib}

\singlespace

\shortauthors{Kogut et al.}
\shorttitle{ARCADE 2 Observations of Galactic Radio Emission}
\slugcomment{Draft December 15, 2008}

\begin{document}

\title{ARCADE 2 Observations of Galactic Radio Emission}

\author{A. Kogut\altaffilmark{1}, 
D. J. Fixsen\altaffilmark{1,2},
S. M. Levin\altaffilmark{3},
M. Limon\altaffilmark{4},
P. M. Lubin\altaffilmark{5},
P. Mirel\altaffilmark{1,6},
M. Seiffert\altaffilmark{4},
J. Singal\altaffilmark{7},
T. Villela\altaffilmark{8},
E. Wollack\altaffilmark{1},
C. A. Wuensche\altaffilmark{8}
}

\altaffiltext{1}{Code 665, Goddard Space Flight Center, Greenbelt, MD 20771}
\altaffiltext{2}{University of Maryland}
\altaffiltext{3}{Jet Propulsion Laboratory, 
California Institute of Technology,
4800 Oak Grove Drive, Pasadena, CA 91109}
\altaffiltext{4}{Columbia Astrophysics Laboratory, 
550W 120th St., Mail Code 5247, New York, NY 10027-6902}
\altaffiltext{5}{University of California at Santa Barbara}
\altaffiltext{6}{Wyle Information Systems}
\altaffiltext{7}{Kavli Institute for Particle Astrophysics and Cosmology, 
SLAC National Accelerator Laboratory, Menlo Park, CA 94025}
\altaffiltext{8}{Instituto Nacional de Pesquisas Espaciais,
Divis\~ao de Astrof\'{\i}sica,
Caixa Postal 515, 12245-970 -  S\~ao Jos\'e dos Campos, SP, Brazil}

\email{Alan.J.Kogut@nasa.gov}

\begin{abstract}
We use absolutely calibrated data from the ARCADE 2 
flight in July 2006
to model Galactic emission at frequencies 3, 8, and 10 GHz.
The spatial structure in the data
is consistent with a superposition of
free-free and synchrotron emission.
Emission with spatial morphology traced by the Haslam 408 MHz survey
has spectral index $\beta_{\rm synch} = -2.5 \pm 0.1$,
with free-free emission
contributing $0.10 \pm 0.01$ of the total Galactic plane emission 
in the lowest ARCADE 2 band at 3.15 GHz.
We estimate the total Galactic emission
toward the polar caps
using either 
a simple plane-parallel model with $\csc(|b|)$ dependence
or
a model of high-latitude radio emission traced by the 
COBE/FIRAS map of C{\sc ii} emission.
Both methods are consistent with a single power-law
over the frequency range 22 MHz to 10 GHz,
with total Galactic emission towards the north polar cap
$T_{\rm Gal} = 0.498 \pm 0.028$ K
and
spectral index $\beta = -2.55 \pm  0.03$
at reference frequency 1 GHz.
The well calibrated ARCADE 2 maps
provide a new test for spinning dust emission,
based on the integrated intensity 
of emission from the Galactic plane
instead of cross-correlations with the thermal dust spatial morphology.
The Galactic plane intensity measured by ARCADE 2
is fainter than predicted by models
without spinning dust,
and is consistent with spinning dust
contributing $0.4 \pm 0.1$ of the Galactic plane
emission at 22 GHz.

\end{abstract}

\keywords{
radio continuum: ISM,
radiation mechanisms: non-thermal,
cosmic microwave background
}



\section{INTRODUCTION}

The cosmic microwave background (CMB)
is a valuable probe of physical conditions
in the early universe.
Its frequency spectrum records the history
of energy transfer
between the evolving matter and radiation fields
to constrain the energetics of the early universe.
We view the CMB through 
diffuse emission from the interstellar medium.
At centimeter wavelengths,
the dominant contributions
are from synchrotron emission
originating from electrons accelerated in the Galactic magnetic field,
and free-free emission (thermal bremsstrahlung)
from electron-ion collisions.
Thermal emission from interstellar dust
is negligible,
but electric dipole emission from a population of small,
rapidly rotating dust grains
could contribute a substantial fraction
of the total Galactic emission
at wavelengths near 1 cm
\citep{draine/lazarian:1998,
dobler/finkbeiner:2007,
miville/etal:2008}.
Absolute measurements of the sky temperature at centimeter wavelengths
measure diffuse emission
to separate Galactic from primordial emission
and
provide information on physical processes in the interstellar medium.

\begin{figure*}[t]
\includegraphics[angle=90,width=7.0in]{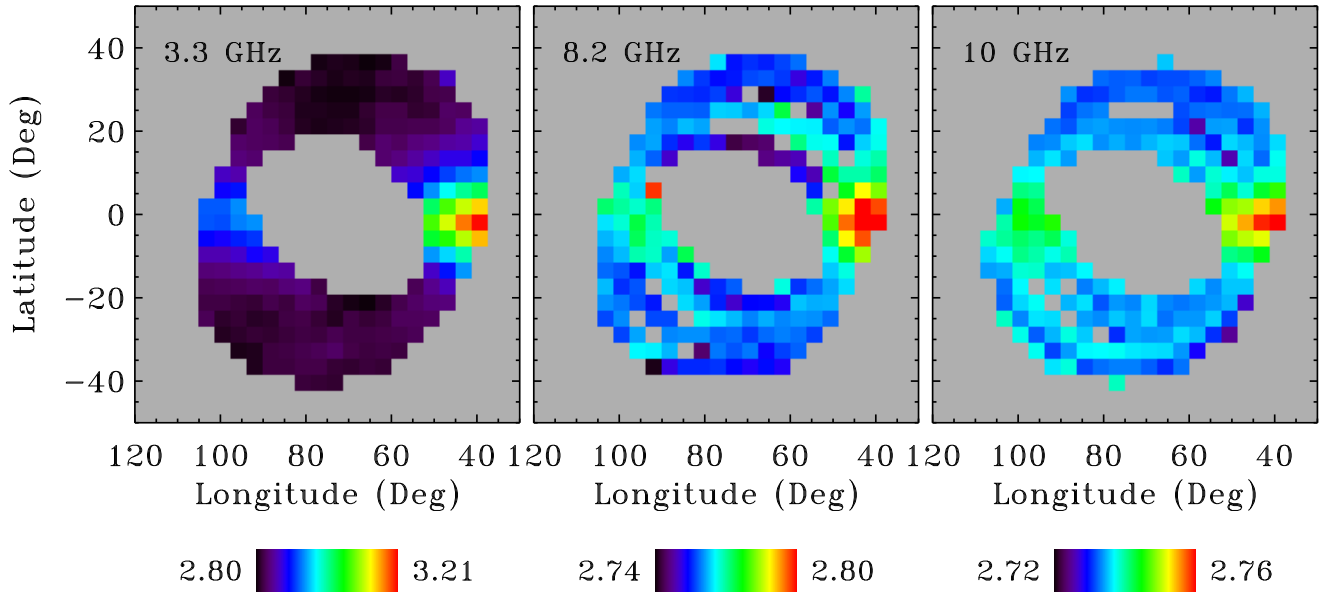}
\caption{
Sky maps from the ARCADE 2 2006 flight
(thermodynamic temperature in Galactic coordinates).
Un-observed regions are masked in grey.
The Galactic plane is clearly visible.
\label{sky_maps} 
}
\end{figure*}

The Absolute Radiometer for Cosmology, Astrophysics, and Diffuse Emission
(ARCADE)
is an instrument to measure the absolute temperature of the sky
in search of distortions from a blackbody spectrum.
ARCADE operates at centimeter wavelengths
between
full-sky surveys at radio frequencies below 3 GHz
and the Far Infrared Absolute Spectrophotometer (FIRAS)
survey at frequencies above 60 GHz.
It consists of a set of cryogenic radiometers
observing at 37 km altitude from a balloon payload.
Each radiometer uses a double-nulled design,
measuring the temperature difference
between a corrugated horn antenna
and an internal reference
as the antenna alternately views
the sky or a full-aperture blackbody calibrator.
The internal reference can be adjusted
to null the antenna-reference signal difference,
while the calibrator temperature can be independently adjusted 
to null the sky-calibrator signal difference.
ARCADE thus measures small shifts about a precise blackbody,
greatly reducing dependence on instrument calibration and stability.
The calibrator, antennas, internal reference, 
and radiometer front-end amplifiers
are mounted within a large liquid helium Dewar
and are maintained near thermal equilibrium with the CMB.
Boiloff helium vapor,
vented through the aperture,
forms a barrier between the instrument and the atmosphere
to allow operation in full cryogenic mode
with no windows between the optics and the sky.

A 2-channel prototype (ARCADE 1)
flew in 2001 and again in 2003,
observing the sky at 10 and 30 GHz
to demonstrate the feasibility 
of cryogenic open-aperture optics
\citep{kogut/etal:2004, fixsen/etal:2004}.
A second-generation 6-channel instrument (ARCADE 2)
flew in 2005 and again in 2006.
A motor failure in 2005
allowed only a single channel to view the sky
\citep{singal/etal:2006}.
The 2006 flight successfully obtained observations
at 3.3, 8.3, 10.2, 30, and 90 GHz.
Data from the 2006 flight are consistent with
a blackbody CMB spectrum,
but show a clear detection
of an extragalactic radio background.
This paper describes the analysis of Galactic emission
from the 2006 flight.
Companion papers
describe the ARCADE 2 instrument 
\citep{singal/etal:2008},
the calibration and sky temperature analysis
\citep{fixsen/etal:2008}
and the implications 
of the detected extra-Galactic radio background
\citep{seiffert/etal:2008}.

\section{Observations and Sky Maps}

The ARCADE 2 instrument launched from 
the Columbia Scientific Balloon Facility in Palestine, TX
carrying a complement
of 7 Dicke-switched radiometers
and 1800 liters of liquid helium.
During ascent,
the Dicke switch failed
in one radiometer.
The remaining radiometers
began sky observations
on 2006 July 22
at 05:08 UT,
continuing 
through 08:11 UT
just prior to flight termination.

Figure \ref{sky_maps} shows the absolute sky temperatures
binned by Galactic coordinates.
Each radiometer views the sky
through a corrugated conical feed horn,
scaled with wavelength
to produce a Gaussian beam
with 11\fdg6 ~full width at half maximum
\citep{singal/etal:2005}.
The beams point 30\arcdeg ~from the zenith
and the entire payload spins at 0.6 RPM
so that the beams
scan a circle of 60\arcdeg ~diameter
centered on the zenith.
The spin frequency is comparable to the
$1/f$ knee of the instrument noise,
so that striping at the mK level
is visible in the binned data.
Each radiometer observed roughly 7\% of the sky.

\begin{table}[b]
\begin{center}
\caption{Frequency Bands For Foreground Analysis}
\label{rad_table}
\begin{tabular}{c c c}
\tableline
\tableline
Frequency & Bandwidth & White Noise 	\\
(GHz)     & (MHz)     & mK s$^{1/2}$ 	\\
\tableline
3.15  & 210 & 11.8 	\\
3.41  & 220 & 10.1 	\\
7.98  & 350 & 5.5	\\
8.33  & 350 & 5.2 	\\
9.72  & 860 & 3.0 	\\
10.49 & 860 & 3.0 	\\
\tableline
\end{tabular}
\end{center}
\end{table}

The measured sky temperatures
are dominated by the CMB monopole.
Galactic emission is clearly visible
in the data at 3, 8, and 10 GHz.
At higher frequencies,
the smaller Galactic signal
and higher instrument noise
combine to prevent a clear detection of Galactic emission.
Throughout this paper,
we consider only data from the
3, 8, and 10 GHz radiometers.
Table \ref{rad_table}
summarizes the in-flight performance
of each radiometer.
We compute the instrument noise by comparing the
variance of the data within each sky bin
to the number of observations in that bin.
This value for the noise
thus includes calibration
as well as the effect of low-frequency striping.

The observed emission is a combination of 
CMB, synchrotron, and free-free emission.
Synchrotron emission
results from the acceleration of cosmic ray electrons
in the Galactic magnetic field.
For a power-law distribution of electron energies
$N(E) \propto E^{-p}$
propagating in a uniform magnetic field,
the synchrotron emission is also a power law,
\begin{equation}
T_A(\nu) \propto \nu^{-\beta_s}
\label{synch_power_law}
\end{equation}
where
$T_A$ is antenna temperature,
$\nu$ is the radiation frequency,
and
$\beta_s = -(p+3)/2$
\citep{rybicki/lightman:1979}.
Free-free emission is also a power law
in antenna temperature,
with spectral index
$\beta_{\rm ff} = -2.15$.

\begin{figure}[t]
\plotone{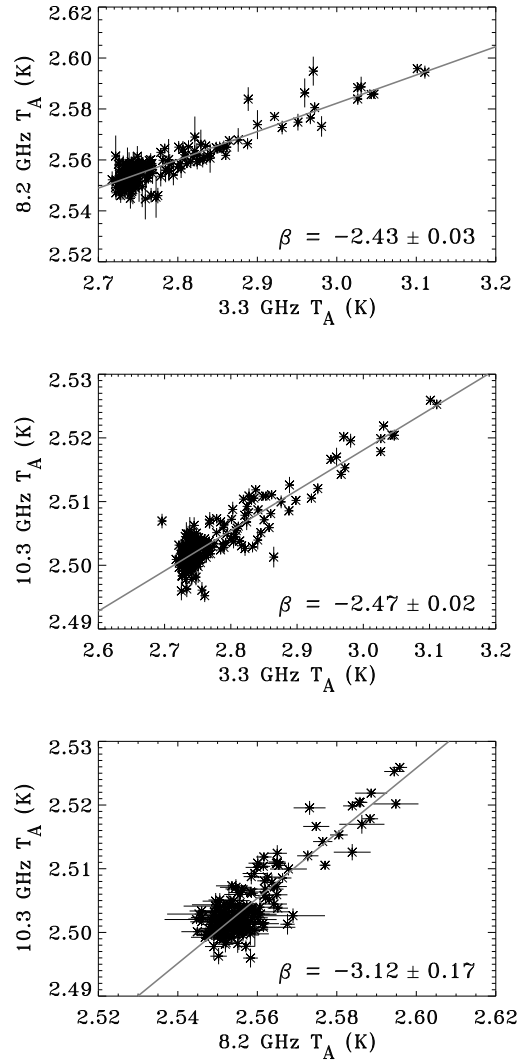}
\caption{
TT plots from the ARCADE 2 sky maps
in units of antenna temperature.
The best-fit spectral index (solid line)
is consistent with a superposition 
of synchrotron and free-free emission.
\label{tt_fig} 
}
\end{figure}

We evaluate the observed spectral index
in the ARCADE 2 sky maps
by plotting the temperature observed at one frequency
against the temperature of the same sky pixel
observed at a different frequency.
Since the ARCADE 2 absolute calibration
is derived from comparison of sky data
to a blackbody calibrator of known physical temperature,
the sky solutions
have units of thermodynamic temperature
\citep{fixsen/etal:2008}.
In order to evaluate the spectral index,
we convert the sky maps
(Fig. \ref{sky_maps})
from thermodynamic to antenna temperature
\begin{equation}
T_A = \frac{x}{\exp(x) - 1} T
\label{tant_eq}
\end{equation}
where
$T$ is thermodynamic temperature,
$x = h \nu / k T$,
$h$ is Planck's constant,
and
$k$ is Boltzmann's constant.
Figure \ref{tt_fig} 
shows the resulting TT plots.

The same Galactic features are observed at each frequency.
A linear fit to each plot
yields spectral index
$\beta = -2.43 \pm 0.03$
from 3.3 to 8.3 GHz,
$\beta = -2.47 \pm 0.02$
from 3.3 to 10.3 GHz,
and
$\beta = -3.12 \pm 0.17$
from 8.3 to 10.3 GHz,
where the quoted errors
include the measurement uncertainties
in both maps for each plot.
The measured index
is consistent with a superposition of
synchrotron and free-free emission.

\section{Spatial Structure}

A search for distortions in the CMB spectrum
requires correction of Galactic foreground emission.
The ARCADE 2 data
has only six frequency channels
containing significant Galactic emission
(Table \ref{rad_table}),
and can not uniquely separate the observed emission
into individual components
(synchrotron, free-free, or other sources).
We combine the ARCADE 2 data with other sky surveys
in order to model foreground emission.

A widely-used technique
describes the observed emission at each frequency $\nu$
as a linear combination of fixed ``template'' maps,
\begin{equation}
T_A(\nu,p) = \sum_i \alpha_i(\nu) G_i(p)
\label{template_eq}
\end{equation}
where
$p$ is a pixel index
and $G_i(p)$ are a set of sky maps
tracing different components of the interstellar medium.
The spectral dependence of the fitted coefficients $\alpha(\nu)$
can then be ascribed to the component traced by each template.

Several choices for template maps tracing free-free emission are available.
\citet{gold/etal:2008} use a maximum-entropy algorithm
to derive a map of free-free emission
based on the
Wilkinson Microwave Anisotropy Probe (WMAP)
5-year data
and a compilation of H$\alpha$ emission
\citep{finkbeiner:2003}.
Since the maximum-entropy method (MEM) template
is heavily weighted by the microwave data
in the Galactic plane,
it minimizes extinction effects in the H$\alpha$ map.
As alternatives to the maximum-entropy template,
we also use either
a map of thermal dust emission
(\citet{fds:1999} model 8)
or the FIRAS map of C{\sc ii} emission at 158 $\mu$m
\citep{fixsen/etal:1999},
neither of which is affected by extinction.

The choice of synchrotron template depends on 
the treatment of spatial variation in the synchrotron spectral index.
To the extent that this variation can be neglected,
the 408 MHz survey
\citep{haslam/etal:1981}
provides a high signal-to-noise ratio tracer for synchrotron emission.
Since the 408 MHz survey
contains both free-free and synchrotron emission,
we correct it for free-free emission
using the MEM free-free template
scaled to 408 MHz using a spectral index $\beta_{\rm ff} = -2.15$
so that to first order
the template represents only synchrotron emission.
We convolve both the synchrotron and free-free templates
to match the ARCADE 11\fdg6 ~beam width.

The ARCADE 2 data are dominated by the CMB monopole,
even towards the Galactic plane.
Both the Galactic and extragalactic signals
also contain monopole terms,
as do the synchrotron and free-free templates.
We thus include a monopole template as a nuisance parameter
and compute the template coefficients $\alpha_i(\nu)$
by minimizing
\begin{equation}
\chi^2 = \sum_p 
	\frac{ \left[ T_A(p, \nu) - \alpha_0(\nu) 
			     - \alpha_{\rm ff}(\nu) G_{\rm ff}(p)
			     - \alpha_{\rm s}(\nu)  G_{\rm s}(p) 
		\right]^2 }
	     { \sigma(p, \nu)^2 }
\label{template_chi2_eq}
\end{equation}
where
$\alpha_0$, $\alpha_{\rm ff}$, and $\alpha_s$
are the coefficients for the
monopole, free-free, and synchrotron templates, respectively,
and
$\sigma(p, \nu)$ is the instrument noise
in each pixel and frequency channel.

The spatial structure in the ARCADE 2 data may be described
as a superposition of emission
traced by the Haslam 408 MHz survey
and
the FIRAS map of C{\sc ii} emission.
Table \ref{coeff_table} shows the fitted coefficients
for the best-fit template combination.
Fitting the coefficients from each template
to a power law in frequency
yields spectral indices
$-2.5 \pm 0.1$ for emission traced by the 408 MHz survey
and
$-2.0 \pm 0.1$ for emission traced by the C{\sc ii} map.

\begin{table}[t]
\begin{center}
\caption{Coefficients for Spatial Templates}
\label{coeff_table}
\begin{tabular}{l c c}
\tableline
\tableline
Frequency & $\alpha_{\rm ff}$	   & $\alpha_s$ \\
(GHz)	  & (mK nW$^{-1}$ m$^2$ sr)  & (mK/K) \\   
\tableline
3.15  &   3.22 $\pm$ 0.11  &   2.02 $\pm$ 0.05	\\
3.41  &   3.06 $\pm$ 0.09  &   1.70 $\pm$ 0.04	\\
7.98  &	  0.26 $\pm$ 0.05  &   0.24 $\pm$ 0.02  \\
8.33  &   0.28 $\pm$ 0.05  &   0.24 $\pm$ 0.02	\\
9.72  &   0.39 $\pm$ 0.03  &   0.04 $\pm$ 0.01	\\
10.49 &   0.37 $\pm$ 0.03  &   0.05 $\pm$ 0.01	\\
\tableline
\end{tabular}
\end{center}
\end{table}

The best template model
uses the 408 MHz survey to trace synchrotron emission
and the FIRAS map of C{\sc ii} line emission
to trace free-free emission.
We have repeated the template fits
using different tracers for synchrotron or free-free emission,
obtaining broadly similar results.
In all cases, the best choice for free-free template
is either the C{\sc ii} map or the thermal dust map.
We note that spatial structure in the ARCADE 2 maps is dominated
by the Galactic plane,
where extinction corrections
for free-free maps derived from H$\alpha$ emission are worst.
The ratio of emission 
traced by the C{\sc ii} free-free template
to emission
traced by the 408 MHz synchrotron template is
$\langle T_{\rm ff} / T_{\rm synch} \rangle = 0.16 \pm 0.01$
~evaluated for latitudes $|b| < 10\arcdeg$
~in the lowest ARCADE 2 channel at 3.15 GHz.
The ratio of free-free emission
to the total Galactic plane emission is
$\langle T_{\rm ff} / T_{\rm total} \rangle = 0.10 \pm 0.01$.

\section{Galactic Polar Cap Temperature}

The sky temperature measured by ARCADE 2 includes contributions from
Galactic, extragalactic, and cosmic sources.  
The 2.7 K cosmic microwave background dominates the measured temperatures.
Galactic emission is dominated by synchrotron emission,
with a smaller contribution from free-free sources.  
The integrated contribution of similar synchrotron
and free-free emission in external galaxies constitutes 
an isotropic extragalactic radio background.  
Accurate measurement of the extragalactic or cosmic backgrounds
requires reliable determination of both the spatial structure
and the zero level of the Galactic emission.

\begin{figure}[b]
\plotone{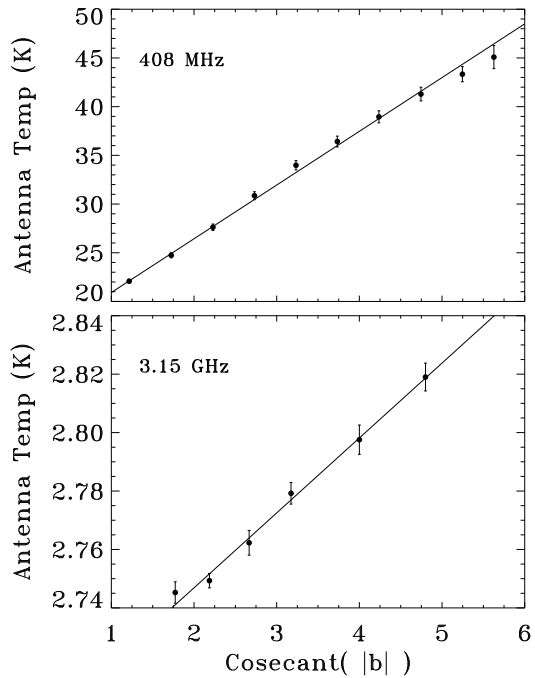}
\caption{
Temperature of the 408 MHz survey (top)
and ARCADE 3.15 GHz channel (bottom)
binned by $\csc(b)$ for latitudes $b > 10\arcdeg$
(Northern hemisphere).
Sky temperatures include the CMB and any extragalactic background.
The scatter about the best-fit line 
results from the higher-order spatial structure in the maps.
For clarity, the plotted uncertainties
have been inflated
by a factor of 20 (408 MHz survey)
or 5 (ARCADE);
the actual statistical uncertainties for the mean of each bin
are smaller than the symbol size.
The data are consistent with a plane-parallel model 
whose measured slope provides an estimate of
the Galactic emission at the poles.
\label{csc_fig} 
}
\end{figure}

Although the cosmic microwave background
may be distinguished from
Galactic or extragalactic radio emission
by the different frequency dependences,
spectral fitting alone
can not distinguish Galactic emission
from an extragalactic component of similar spectral behavior.
The template model above reproduces the spatial structure 
of the observed Galactic emission,
but is insensitive to emission described by an additive constant in each map.
If the total Galactic brightness 
were determined along some fiducial line of sight,
we could add a monopole term to the template model
to match the total Galactic brightness along that line of sight.
The adjusted model would then characterize Galactic emission
along all other lines of sight.

The north and south polar caps (latitude $|b| > 75\arcdeg$)
provide convenient reference lines of sight.
We use two independent methods to estimate 
the total Galactic emission toward the Galactic polar caps.
The first method
relies solely on the spatial morphology,
assuming a plane-parallel model
to estimate the total Galactic emission at the polar caps.
A second, independent method
compares the spatial morphology of the ARCADE 2 and other radio maps
to template maps of atomic line emission
in order to estimate the 
amount of radio emission
per unit line emission.
Since line emission can unambiguously be attributed to the Galaxy,
the total Galactic radio emission toward the polar caps
may then be estimated by
scaling the total line emission at the caps
by the observed ratio of radio to line emission.
Both methods yield similar values for the 
total Galactic emission in the polar cap regions.

\subsection{Polar Cap Temperature From $\csc(b)$}

A commonly used method to estimate total Galactic brightness
models the Galaxy as a simple plane-parallel structure
and fits the spatial distribution of the radio continuum
at Galactic latitude $|b| > 10\arcdeg$
~to the form
\begin{equation}
T_A(\nu, p) = c(\nu) + T_G(\nu) \csc( |b| )
\label{csc_eq}
\end{equation}
where $b$ is the Galactic latitude of each pixel $p$
and the constant $c(\nu)$ accounts for
the contribution from the CMB or extragalactic backgrounds
at each frequency $\nu$.
Figure \ref{csc_fig} shows typical results for the northern hemisphere,
both for the lowest frequency ARCADE 2 channel
and the 408 MHz survey.
The high-latitude structure of the radio sky 
is consistent with a plane-parallel model,
whose slope defines the antenna temperature $T_G(\nu)$ 
of Galactic emission toward the Galactic poles.

\begin{figure}[t]
\includegraphics[angle=90,width=3.5in]{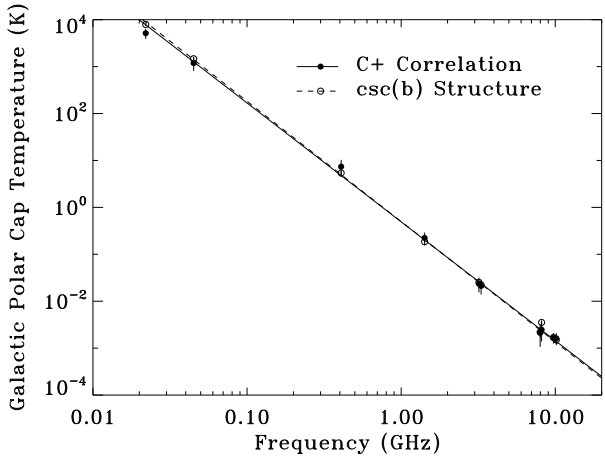}
\caption{
Galactic emission $T_G(\nu)$
towards the North galactic pole,
derived from 
a cosecant fit to the spatial structure (open circles)
and the radio/C{\sc ii} correlation (filled circles).
Both methods agree and are consistent with a single power law
over the frequency range 22 MHz to 10 GHz.
The fitted Galactic temperature towards the polar caps
can be combined with the template model of the spatial structure
to fully specify the Galactic model.
\label{cap_model_vs_freq} 
}
\end{figure}

We repeat the cosecant fit (Eq. \ref{csc_eq})
independently for the northern and southern hemispheres,
using each of the ARCADE 2 sky maps
as well as a selection of 
lower-frequency radio surveys.
Surveys at 22 MHz
\citep{roger/etal:1999},
45 MHz 
\citep{maeda/etal:1999,alvarez/etal:1997},
408 MHz
\citep{haslam/etal:1981},
and 1420 MHz
\citep{reich/etal:2001,reich/reich:1986}
have full or nearly full sky coverage
at frequencies where the sky brightness is dominated
by Galactic radio emission.
For each sky map,
we compute the cosecant slope $T_G(\nu)$
and its uncertainty,
including both statistical and calibration uncertainties.
Although the binned data are broadly consistent
with a plane-parallel model,
higher-order structure causes
the binned data to show greater scatter about the best-fit slope
than would be expected given the formal statistical uncertainty
in the mean of each bin.
We account for higher-order structure 
by inflating the statistical uncertainty
to force $\chi^2$ to unity per degree of freedom
about the best-fit line.
We then add the calibration uncertainty
(of order 10\% for the low-frequency radio surveys)
in quadrature with the (inflated) statistical uncertainty
to derive the total uncertainty in the cosecant slope
for each sky survevy.

Figure \ref{cap_model_vs_freq}
shows the Galactic temperature $T_G(\nu)$ 
towards the north polar cap
derived from the spatial morphology of the maps.
Although the sky has significant longitudinal substructure
within each latitude bin,
the binned data for each map
strongly supports the gross morphology
dominated by a plane-parallel structure.
The derived polar cap temperature
is consistent with a single power law
over the frequency range 22 MHz to 10 GHz.

\begin{figure}[b]
\includegraphics[angle=90,width=3.5in]{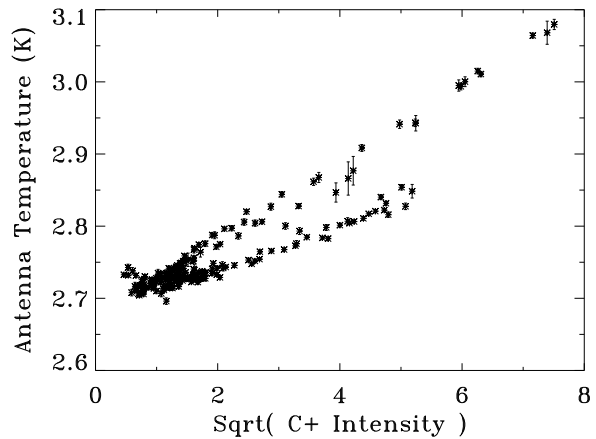}
\caption{
Correlation between the ARCADE 3.15 GHz intensity
and the C{\sc ii} 158 $\mu$m atomic line
in units nW m$^{-2}$ sr$^{-1}$.
The data clearly divide into two regions 
each with radio intensity proportional to the square root
of the C{\sc ii} intensity.
The upper track originates from pixels near the Galactic center,
while the lower track originates from pixels near the Cygnus region.
\label{arc_cii_tt} 
}
\end{figure}

\subsection{Radio/C{\sc ii} Correlation}

Maps of line emission provide a valuable tracer of Galactic emission.
Line emission from the diffuse interstellar medium 
can be measured over the full sky,
has a well-defined amplitude along every line of sight,
and does not suffer from extragalactic contamination.  
Several lines could be used to trace Galactic microwave emission.
H$\alpha$ emission from 3--2 transition in neutral atomic hydrogen
has been mapped over the full sky 
and is a well-established tracer of free-free emission.  
However, H$\alpha$ emission predominantly traces
emission from the warm ionized medium,
and may not accurately trace emission from 
the brighter synchrotron component.
In addition, H$\alpha$ emission suffers from dust extinction 
in the Galactic plane 
making it less reliable in regions where the Galactic radio emission 
is brightest.
The 21 cm fine-structure line from neutral hydrogen
has also been mapped over the full sky.
However, since HI emission originates from 
the neutral component of the interstellar medium,
it is unlikely to trace microwave emission
from ionized regions.
The C{\sc ii} line at 158 $\mu$m wavelength 
from singly ionized carbon 
can be used to trace diffuse radio emission.  
It is optically thin even in the plane,
does not suffer from dust extinction, 
and is an important cooling mechanism for the diffuse interstellar medium.  
It has been mapped over the full sky by the COBE/FIRAS instrument
\citep{fixsen/etal:1999}.

Figure \ref{arc_cii_tt}
shows the correlation between the ARCADE 3.15 GHz map
and the C{\sc ii} 158 $\mu$m map.
The C{\sc ii} map has been smoothed to angular resolution 11\fdg6
~to match the ARCADE resolution.
The data segregate into two tracks,
an upper track from emission near the Galactic center
and a lower track from emission near the Cygnus region.
Both tracks show radio emission proportional to the
square root of the C{\sc ii} intensity.
This would be expected if the radio emission 
were proportional to the density $n$ in the interstellar medium
(e.g. from synchrotron emission),
while the collisionally excited C{\sc ii} intensity
were proportional to $n^2$.
Microwave free-free emission,
proportional to $n^2$,
would result in a curved track in Figure \ref{arc_cii_tt}.
The absence of any such curvature
provides additional evidence that free-free emission
is faint compared to synchrotron
at the ARCADE 2 bands.
Note, though,
that if a separate template is provided for the synchrotron component,
the C{\sc ii} map
may be used
to trace the fainter free-free component
(at least at the ARCADE angular resolution).

\begin{figure}[t]
\includegraphics[angle=90,width=3.5in]{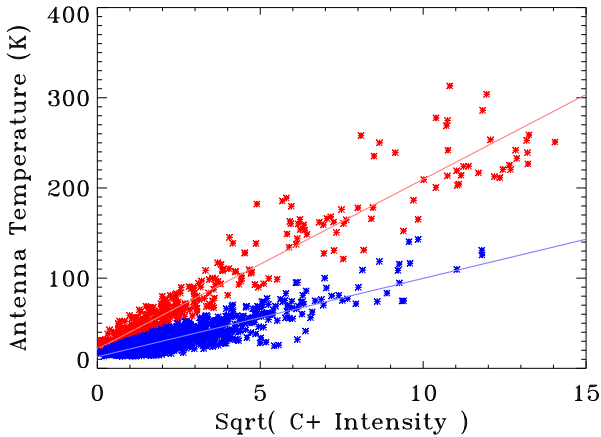}
\caption{
Correlation between the full-sky 408 MHz survey
and the C{\sc ii} 158 $\mu$m atomic line
in units nW m$^{-2}$ sr$^{-1}$.
The bifurcation observed in the limited ARCADE 2 sky coverage
persists over larger regions of the sky.
We fit the data to two lines
and iteratively assign each point to one line or the other
(see text for details).
Colors indicate the final assignments.
\label{has_vs_cii} 
}
\end{figure}

The prominent segregation in Fig \ref{arc_cii_tt} 
results at least in part from the limited sky coverage.
The brightest regions in the ARCADE 2 maps
occur at the edge of the Galactic center region
and again at the edge of the Cygnus region.
The ratio of C{\sc ii} intensity to the far-infrared continuum
is known to differ by a factor of two between these regions
\citep{fixsen/etal:1999}.
We investigate the extent to which the ratio of C{\sc ii} intensity 
to radio continuum emission varies
by performing a similar analysis 
on radio surveys
at 22 , 45, 408, and 1420 MHz.
All exhibit evidence for similar segregation,
although not as pronounced as for the smaller ARCADE 2 sky coverage.

\begin{table*}[t]
\begin{center}
\caption{Radio/C{\sc ii} Correlation Slope}
\label{correlation_table}
\begin{tabular}{l c c c c c}
\tableline
\tableline
Frequency & \multicolumn{5}{c}{Correlation Slope $a$ (mK [ nW m$^{-2}$ sr$^{-1}$]$^{-0.5}$ )} \\
(GHz)	  &  Mask 1 & ~~~ & Mask 2 & ~~~ & Mean \\
\tableline
 0.022 	& $(1.12 \pm 0.06) \times 10^4$ & 
 	& $(0.74 \pm 0.02) \times 10^4$ & 
	& $(0.93 \pm 0.19) \times 10^4$		\\
 0.046  & $(2.74 \pm 0.07) \times 10^3$ & 
 	& $(1.56 \pm 0.03) \times 10^3$ & 
	& $(2.15 \pm 0.63) \times 10^3$ 	\\
 0.408  & $18.1 \pm 0.3$   & 
 	& $8.5  \pm 0.1$   & 
	& $13.3 \pm 5.0$ 			\\
 1.420  & $(5.11 \pm 0.08) \times 10^{-1}$ & 
 	& $(2.94 \pm 0.04) \times 10^{-1}$ & 
	& $(4.02 \pm 1.10) \times 10^{-1}$	\\
 3.195  & $(5.63 \pm 0.12) \times 10^{-2}$ & 
 	& $(2.88 \pm 0.09) \times 10^{-2}$ & 
	& $(4.25 \pm 1.38) \times 10^{-2}$	\\
 3.300  & $(5.05 \pm 0.11) \times 10^{-2}$ & 
 	& $(2.61 \pm 0.08) \times 10^{-2}$ & 
	& $(3.82 \pm 1.20) \times 10^{-2}$	\\
 8.15   & $(5.73 \pm 0.37) \times 10^{-3}$ &	
 	& $(1.99 \pm 0.36) \times 10^{-3}$ &
	& $(3.86 \pm 1.87) \times 10^{-3}$	\\
 8.33   & $(6.32 \pm 0.34) \times 10^{-3}$ & 
 	& $(2.59 \pm 0.26) \times 10^{-3}$ & 
	& $(4.46 \pm 1.86) \times 10^{-3}$	\\
 9.72   & $(3.68 \pm 0.15) \times 10^{-3}$ & 
 	& $(2.37 \pm 0.16) \times 10^{-3}$ & 
	& $(3.03 \pm 0.66) \times 10^{-3}$	\\
10.15   & $(3.67 \pm 0.16) \times 10^{-3}$ & 
	& $(2.17 \pm 0.18) \times 10^{-3}$ & 
	& $(2.92 \pm 0.75) \times 10^{-3}$	\\
\tableline
\end{tabular}
\end{center}
\end{table*}

Figure \ref{has_vs_cii} shows the correlation
between the C{\sc ii} map and the 408 MHz survey,
after smoothing the 408 MHz survey with a 5\arcdeg ~FWHM Gaussian 
to approximate the 7\arcdeg ~tophat FIRAS beam.
We quantify the tendency for the data to segregate into two tracks
by fitting two lines to the correlation plot.
After a first guess at the fit parameters
(intercept and slope for each line),
we assign each point to one line or the other
based on the difference in antenna temperature
between that point and either line.
We then re-compute the fit parameters for each line
using only those points assigned to that line,
and then re-assign all points based on the new fit parameters.
Several iterations suffice to produce a stable solution
independent of the initial guess.
The segregation is statistically significant:
fitting the 2538 points to two lines
reduces the $\chi^2$ by a factor of 4 
compared to a single line fit to all points.

\begin{figure}[t]
\plotone{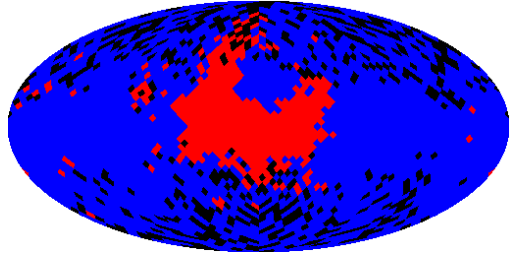}
\caption{
Pixel masks in Galactic coordinates showing the location of
emission components from Fig. \ref{has_vs_cii}.
Pixels with higher radio/C{\sc ii} correlation
(red) lie near the Galactic center
or along the North Galactic spur (radio Loop I).
Pixels in blue have radio/C{\sc ii} correlation
smaller by a factor of roughly 2.
Pixels in black have poor signal to noise ratio 
in the C{\sc ii} map and are not used.
\label{has_mask} 
}
\end{figure}

The segregation in the correlation plot
can be related to known structures in the radio sky.
The points in Fig. \ref{has_vs_cii}
are colored to highlight the segregation into two components.
Figure \ref{has_mask} maps each point in Galactic coordinates.
Points associated with higher radio/C{\sc ii} ratio
lie toward the Galactic center 
and the North Galactic Spur (radio Loop I).
Both are regions with enhanced synchrotron emission.

\subsection{Polar Cap Temperature From C{\sc ii} Correlation}

The observed correlation between radio emission and C{\sc ii} intensity
allows a well-defined determination of the 
associated Galactic radio emission
toward the Galactic poles.
We use the pixel masks 
defined by the 408 MHz/C{\sc ii} correlation
(Figure \ref{has_mask})
to fit each radio survey to the form
\begin{equation}
T_A(\nu, p) = \sum_{i=1}^2 b_i(\nu) + a_i(\nu) (~I_C(p)~)^{0.5},
\label{correlation_eq}
\end{equation}
where
$T_A$ is the antenna temperature at frequency $\nu$,
$I_C$ is the C{\sc ii} intensity,
and
$p$ denotes the pixels within 
one of the two spatial masks.

The intercepts $b_i$ for the two masks
include contributions from
extragalactic sources 
(including, notably, the CMB monopole)
and can not uniquely determine absolute level of Galactic emission.
We instead define the Galactic emission at the polar caps
by extrapolating the observed radio/C{\sc ii} correlation
to high latitude:
\begin{equation}
T_G(\nu) = a(\nu) ~I_{\rm cap}^{0.5},
\label{zero_def}
\end{equation}
where
$I_{\rm cap}$
is the mean intensity of the C{\sc ii} map
at the polar caps (Galactic latitude $|b| > 75\arcdeg$).

The two spatial masks separate the sky 
into regions with higher or lower radio emission
per unit C{\sc ii} intensity.
The difference between these two components
is less clear at high latitude
where both the radio emission and C{\sc ii} intensity become weaker.
The spatial segregation of the two components
suggests that the south polar cap region
is associated with the weaker radio/C{\sc ii} component
(Figure \ref{has_mask}).
The brighter radio regions associated 
with the North Galactic spur, however, 
extend to high latitude
so that the mean radio brightness
associated with C{\sc ii} emission
toward the north polar cap region
probably lies between the limiting cases
established by the two masks.
We therefore use the mean of the slopes from the two masks
to estimate the Galactic radio emission at the north polar cap,
with uncertainty broad enough to bracket the two cases.
The mean value thus derived is consistent
with similar analysis fitting a single slope to the full sky.
Within the restricted ARCADE 2 sky coverage, however,
the value of a single slope fitted to the entire observed region
could vary significantly
as the sky coverage shifts slightly
to include more or less of the Galactic center region.
The mean of the two observed slopes
is less susceptible to variations in the ARCADE 2 sky coverage.

Table \ref{correlation_table}
shows the correlation slopes $a_i(\nu)$
fitted to each mask 
for both the ARCADE 2 data
and lower-frequency sky surveys.
Figure \ref{cap_model_vs_freq} 
compares the Galactic temperature towards the north polar cap,
derived from mean C{\sc ii} correlation,
to the value derived from the cosecant fit.
Both methods provide similar estimates for the Galactic component
of the total brightness temperature towards the north polar cap.
This Galactic component is consistent
with a single power law
\begin{equation}
T_G(\nu) = T_{\rm Gal} (\nu/\nu_0)^\beta
\label{cap_def}
\end{equation}
with spectral index
$\beta = -2.55 \pm  0.03$
and amplitude
$T_{\rm Gal} = 0.498 \pm 0.028$ K
at reference frequency $\nu_0$ = 1 GHz.

The analysis above uses a single map (C{\sc ii} line emission)
to trace Galactic emission from the diffuse interstellar medium.
If a significant fraction of Galactic radio emission
originated from a component of the interstellar medium
not well sampled by C{\sc ii} emission,
the resulting correlation would under-estimate the total Galactic emission
toward the polar cap.
We test for additional microwave emission
from other components of the interstellar medium
by repeating the correlation analysis
using a simultaneous fit to three line maps
chosen to sample different components of the diffuse interstellar medium:
the (square root of the) COBE/FIRAS map of C{\sc ii} emission,
a map of H$\alpha$ emission
\citep{finkbeiner:2003},
and the
Leiden/Argentine/Bonn Galactic HI Survey
\citep{kalberla/etal:2005}.
The simultaneous fit produces results nearly identical
to a fit using just the C{\sc ii} map,
shifting the derived Galactic polar cap temperature
by $11 \pm 21$ mK at reference frequency 1 GHz.
Since the limited ARCADE 2 sky coverage
increases the effect of covariance between the different line maps,
we use the results derived from the single C{\sc ii} correlation.

\begin{table*}[t]
\begin{center}
\caption{Galactic Emission$^a$ Along Selected Lines of Sight}
\label{ref_gal_table}
\begin{tabular}{l l c c c}
\tableline
\tableline
Parameter & Technique & North Polar Cap & South Polar Cap & Coldest Patch \\
\tableline
		  & Radio/C{\sc ii} & $ 0.492 \pm 0.095 $ & $ 0.303 \pm 0.054 $ & $ 0.188 \pm 0.130 $ \\
$T_{\rm Gal}$ (K) & $\csc(|b|)$     & $ 0.499 \pm 0.030 $ & $ 0.366 \pm 0.028 $ & ---		     \\
		  & Weighted Mean   & $ 0.498 \pm 0.028 $ & $ 0.353 \pm 0.025 $ & $ 0.188 \pm 0.130 $ \\
\tableline
		  & Radio/C{\sc ii} & $ -2.53 \pm 0.07  $ & $ -2.59 \pm 0.06  $ & $ -2.56 \pm 0.12  $ \\
$\beta$		  & $\csc(|b|)$	    & $ -2.56 \pm 0.04  $ & $ -2.65 \pm 0.05  $ & ---		     \\
		  & Weighted Mean   & $ -2.55 \pm 0.03  $ & $ -2.63 \pm 0.04  $ & $ -2.56 \pm 0.12  $ \\
\tableline
\end{tabular}
\end{center}

\begin{center}
$^a$ Galactic emission $T_G(\nu) = T_{\rm Gal} (\nu / \nu_0)^\beta$
with reference frequency $\nu_0$ = 1 GHz.
\end{center}
\end{table*}

\subsection{Composite Galactic Model}

Removing Galactic emission from the ARCADE 2 data
requires a model for both the spatial structure
and the total Galactic emission.
At each ARCADE frequency,
we generate a full-sky model of the spatial structure
using full-sky template maps
multipled by the coefficients 
fitted within the region observed by ARCADE 2
(Table \ref{coeff_table}).
We then add a constant to the template model
to match the total Galactic emission
derived from the C{\sc ii} and $\csc(|b|)$ fits
towards selected reference positions.

We independently model the total Galactic emission
toward the north and south Galactic polar caps
($|b| > 75\arcdeg$).
For each reference position,
we compute the total Galactic emission
associated with either 
the plane-parallel structure
(Eq. \ref{csc_eq})
or the C{\sc ii} emission
(Eq. \ref{zero_def}).
We apply each method
to the ARCADE 2 and radio surveys,
and parameterize the resulting multi-frequency results
using a power-law model
(Eq. \ref{cap_def}).
As a cross-check, we also estimate the total Galactic emission
for the coldest patch in the Northern hemisphere,
consisting of all pixels within 15\arcdeg ~of 
$b = 48\arcdeg$,
$l = 196\arcdeg$
(north of the Galactic anti-center).
Since by definition a mid-latitude cold spot
is inconsistent with a plane-parallel structure,
the estimate for this position
is based only on the radio/C{\sc ii} correlation.
Table \ref{ref_gal_table}
shows the estimated Galactic normalization 
$T_{\rm Gal}$
and spectral index $\beta$
for each reference position.
We obtain similar results
using either the spatial morphology
or the radio/C{\sc ii} correlation.

We fix the offset of the template model
(Eq. \ref{template_eq})
by computing the temperature of the template model
towards these same regions,
then adding a constant $\alpha_0(\nu)$
to the template model at each frequency $\nu$
to force the template model to match the power-law model
along these lines of sight.
The resulting composite model of Galactic emission
can be used to correct the calibrated time-ordered data
in order to estimate the CMB monopole temperature
and the extragalactic radio background
\citep{fixsen/etal:2008}.
The offsets derived from the three independent regions
agree within 5 mK at 3 GHz, which we adopt as the
uncertainty in the zero level of composite Galactic model.

The spectral index derived for the Galactic emission
toward the polar caps or coldest region,
$\beta \approx -2.57 \pm 0.03$,
is consistent with the spectral index 
$\beta = -2.5 \pm 0.1$
derived from the synchrotron template fit only to ARCADE 2 data,
indicating that high-latitude Galactic emission
is dominated by synchrotron emission
at frequencies below 10 GHz.
A synchrotron spectral index of -2.57
is significantly flatter than the values 
derived from measurements at higher frequencies,
but is consistent with the majority of measurements
below 10 GHz
(see, e.g., the tabulation in 
\citet{rogers/bowman:2008}).
Notable exceptions are the 
measurements by 
\citet{tartari/etal:2008}
and
\citet{platania/etal:1998},
which prefer $\beta \approx -2.8$
for the synchrotron component.
Both of these results
are derived from a single strip at declination $\delta = +42\arcdeg$
~and may not be representative
of larger regions of the sky.

\section{Spinning Dust}

The ARCADE 2 data can also be used
to search for additional microwave emission
associated with interstellar dust.
The microwave sky is known to contain a component
spatially correlated with far-infrared dust emission
but not with synchrotron-dominated surveys at 408 or 1420 MHz
\citep{kogut/etal:1996a,
kogut/etal:1996b,
oliveira-costa/etal:1997,
leitch/etal:1997,
bennett/etal:2003}.
The frequency spectrum of the correlated component
differs markedly from the emission spectrum of thermal dust,
showing a spectral index $\beta \approx -2.2$
in antenna temperature from 20 to 50 GHz.

Two main candidates have emerged to explain this ``anomalous'' 
correlated emission component.
Electric dipole emission
from a population of small, rapidly rotating dust grains
will produce a spectrum with a broad peak
in the frequency range 20--40 GHz
\citep{draine/lazarian:1998}.
\citet{miville/etal:2008} analyze data from
the Wilkinson Microwave Anisotropy Probe (WMAP) mission
and conclude that spinning dust
accounts for the majority of emission
in the Galactic plane
at 22 GHz.
An alternative model explains the correlated emission
as flat-spectrum synchrotron emission
associated with star formation activity
\citep{bennett/etal:2003,
gold/etal:2008}.
Energy losses as electrons propagate from their origin
steepen the synchrotron index away from these sources.
Since a flat-spectrum component will increasingly dominate
at higher frequencies,
the spatial morphology of synchrotron emission at higher frequencies
should increasingly resemble
thermal dust emission 
from the same star formation activity.

\begin{figure}[b]
\includegraphics[angle=90,width=3.5in]{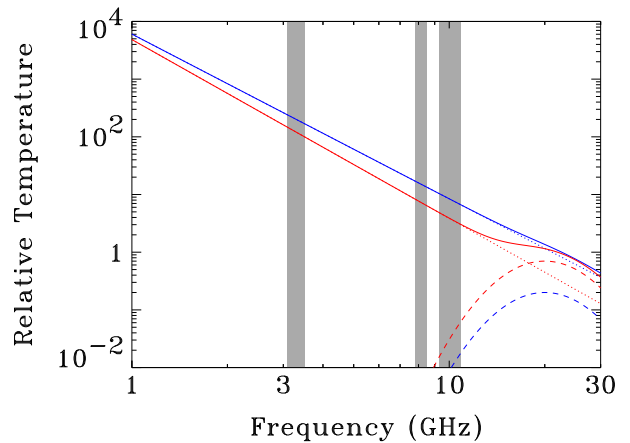}
\caption{
Synchrotron and spinning dust emission
for two models with different spinning dust amplitude.
The red (blue) curves show emission 
from a model with high (low) dust normalization.
Dashed lines show the dust emission,
dotted lines show synchrotron emission,
and solid lines show the combined emission
from each model.
The amplitude of the combined emission
is fixed at both 22 GHz and 408 MHz.
Measurements below 10 GHz are dominated by synchrotron,
which can be used to infer the dust normalization:
models with more dust emission at 22 GHz
have lower synchrotron emission.
Grey bars indicate the ARCADE 2 frequency bands.
\label{spin_model} 
}
\end{figure}

Efforts to distinguish between these models
have largely focused on 
the frequency spectrum
of the dust-correlated component
inferred by 
correlating a map tracing thermal dust emission
against microwave maps at different observing frequencies.
The correlation with thermal dust emission
is seen to peak at frequencies between 20 and 30 GHz, 
falling 
at frequencies below 22 GHz
\citep{oliveira-costa/etal:2004,
fernandez-cerezo/etal:2006,
hildebrandt/etal:2007}.
This falling spectrum at low frequencies 
has been cited as evidence favoring spinning dust models.

The decrease in dust-correlated emission
at frequencies below 20 GHz,
although consistent with spinning dust emission,
does not necessarily support such models
over the flat-spectrum synchrotron alternative.
{\it Both} models predict weaker correlation
with thermal dust emission at frequencies below 20 GHz.
For the spinning dust model,
the weaker correlation
results from the lower amplitude 
of the spinning dust emission.
The flat-spectrum synchrotron model, however, also predicts
a weaker correlation at lower frequencies,
since by construction
the spatial morphology of the sky in this model
varies smoothly from the WMAP data at 22 GHz
(observed to correlate well with the thermal dust morphology)
to the Haslam map at 408 MHz
(observed to correlate only weakly with thermal dust emission).
Model discrimination based on cross-correlations with thermal dust emission
must rely on rigorous statistical comparison of competing models,
and not simply on the general trend toward weaker correlations
at lower frequencies.

A simpler test relies on the absolute spectrum 
of Galactic emission.
At frequencies below 30 GHz,
thermal dust emission is negligible,
so that the total Galactic emission becomes a superposition
of free-free,
synchrotron,
and spinning dust emission.
Both free-free and synchrotron emission
increase monotonically at lower frequencies.
Spinning dust, in contrast,
{\it decreases} in amplitude below 20 GHz,
so that the spectrum of the combined Galactic emission below 20 GHz
can be used to place limits on the contribution
of spinning dust
without reference to detailed spatial correlations 
with a thermal dust template.

Figure \ref{spin_model} illustrates the concept.
Consider Galactic emission consisting
of a superposition of synchrotron and spinning dust
(ignoring for the moment the smaller contribution from
free-free emission).
Following \citet{miville/etal:2008},
we specify the spinning dust amplitude
as a fraction of the total Galactic brightness 
at 22 GHz,
assumed here to represent the peak in the dust spectrum.
As the dust normalization is increased,
the synchrotron amplitude at 22 GHz
must decrease to keep the total emission constant.
The spinning dust spectrum falls rapidly,
so that emission below 10 GHz 
is dominated by the synchrotron component.
A model with high dust normalization
(red curves in Fig. \ref{spin_model})
will thus have fainter synchrotron emission
at frequencies of a few GHz,
while a model with less spinning dust (blue curves)
will have brighter synchrotron emission.
This shift in the amplitude of Galactic emission
at frequencies below 10 GHz
is a sensitive test for spinning dust,
without resort to detailed spatial correlations.

We implement this test using a simple model of Galactic emission.
We assume that the spatial distribution of spinning dust emission
is traced by a template map of thermal dust emission,
and fix the amplitude of associated spinning dust emission
by scaling the template map
so that the re-scaled dust map
forms a specified fraction of the 
Galactic plane intensity measured by the WMAP 22 GHz map.
This allows a simple normalization of spinning dust
in terms of its relative contribution
to the total Galactic plane intensity at 22 GHz.
We use the COBE/DIRBE 240 $\mu$m map of thermal dust emission
\citep{reach/etal:1996}
as the thermal dust template
since it is dominated by thermal dust emission
but unaffected by extinction in the plane.
For specificity, we define 
the Galactic plane mask
using all pixels 
lying within the ARCADE 2 observation pattern
with latitude $|b| < 20\arcdeg$,
and use this mask for all computations
so that the model results
are not affected by differing sky coverage
between ARCADE  2 and other surveys.

\begin{figure}[b]
\plotone{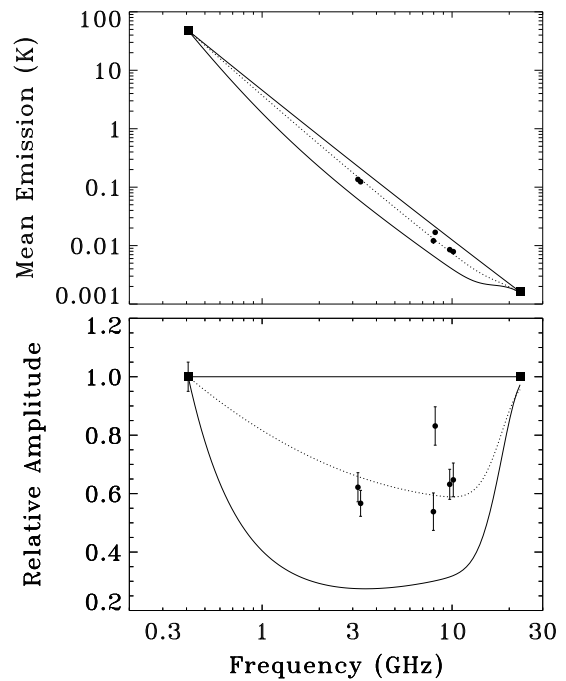}
\caption{
Mean intensity of Galactic plane emission
for the ARCADE 2 data,
compared to model predictions
with and without spinning dust (see text).
Top panel: Model predictions and data.
The upper solid curve shows the 
model prediction for no spinning dust,
averaged over the ARCADE 2 sky coverage with $|b| < 20\arcdeg$.
The lower solid curve shows the model prediction
for spinning dust amplitude equal to 60\% of the
total Galactic plane emission at 22 GHz.
The dotted curve shows the best fit to the ARCADE 2 data.
The spectrum of emission from spinning dust
is assumed to peak at 22 GHz,
so a higher spinning dust fraction at 22 GHz
produces lower total emission at the ARCADE 2 frequencies.
The bottom panel shows the same data,
normalized by dividing each point 
by the model prediction for no spinning dust.
Galactic plane emission observed by ARCADE 2
is consistently fainter
than expected for a model with no spinning dust,
and is consistent
with spinning dust contributing $0.4 \pm 0.1$ 
of the total Galactic plane emission
at reference frequency 22 GHz.
\label{spinning_dust_test} 
}
\end{figure}

We then model the Galactic emission spectrum as follows.
We first correct the WMAP 22 GHz map
and the Haslam 408 MHz map
by subtracting the WMAP maximum-entropy model of free-free emission
\citep{gold/etal:2008}
using a spectral index $\beta_{\rm ff} = -2.15$.
The corrected maps then contain only synchrotron
and (possibly) spinning dust emission.
We assume that the spinning dust spectrum peaks at 22 GHz
and scale the normalized spinning dust map
to lower frequencies
using the
\citet{draine/lazarian:1998} model for the warm neutral medium.
After correction for free-free and spinning dust emission,
the 22 GHz and 408 MHz maps contain only synchrotron emission,
which we use to define the 
synchrotron amplitude and spectral index for each pixel.

The resulting model
(synchrotron, free-free, and spinning dust)
can be used to estimate the combined Galactic emission
at frequencies between 408 MHz and 22 GHz.
We compare the model to the ARCADE 2 data
by smoothing the model map of combined emission
to the ARCADE 2 angular resolution,
and then computing the mean Galactic emission
of the smoothed model
for all pixels within the ARCADE 2 galactic plane mask.
Figure \ref{spinning_dust_test}
shows the result.
If spinning dust is negligible,
the synchrotron contribution at 22 GHz is maximal
and the model approximates the flat-spectrum synchrotron model.
As the spinning dust amplitude increases,
the synchrotron contribution at 22 GHz decreases
and the synchrotron spectral index steepens.
The combination of lower synchrotron amplitude
and falling dust spectrum
combine to lower the total model emission
across the ARCADE 2 frequency bands.
The ARCADE 2 data lie below the model prediction
for no spinning dust,
and are consistent with 
spinning dust contributing $0.4 \pm 0.1$ 
of the total K-band Galactic plane emission.

WMAP is a differential instrument
and is insensitive to any monopole emission component.
The zero level of the 22 GHz map
is set using the cosecant dependence on Galactic latitude
\citep{hinshaw/etal:2008}.
The 408 MHz survey and the ARCADE 2 data,
however, both include monopole contributions.
To prevent discrepant treatment of the map zero levels
from affecting the model predictions,
we remove a monopole from the 408 MHz survey
and the ARCADE 2 sky maps,
and re-set the zero level of each map
using a $\csc(|b|)$ fit to the spatial structure in each map
to match the processing of the WMAP data.
We restrict analysis to pixels at low Galactic latitude 
($|b| < 20\arcdeg$)
where Galactic emission is brightest
so that uncertainties in the zero level have minimal effect.
The largest uncertainty in the analysis
is the correction for free-free emission.
We use the WMAP maximum-entropy model of free-free emission,
derived assuming that spinning dust contributes negligibly
to the total Galactic emission.
We test the sensitivity of the result
to the free-free normalization
by repeating the analysis
using the same maximum-entropy model
scaled by a constant normalization factor.
Changing the free-free normalization by a factor of 2
affects the best-fit spinning dust normalization by approximately 0.08,
and is included in the total quoted uncertainty.

\section{Conclusions}

We use the ARCADE 2 absolutely calibrated observations of the sky
~to model Galactic emission 
at frequencies 3, 8, and 10 GHz
and angular resolution 11\fdg6.
TT plots of the binned sky maps
show net spectral index
$\beta = -2.43 \pm 0.03$ between 3.3 and 8.3 GHz
and
$\beta = -2.47 \pm 0.02$ between 3.3 and 10.3 GHz,
consistent with a superposition of
synchrotron and free-free emission.
The spatial structure in the maps
can be described using two spatial templates,
the Haslam 408 MHz survey
to trace synchrotron emission
and the COBE/FIRAS map of C{\sc ii} emission
to trace free-free emission.
Fitting these templates to the ARCADE 2 maps
yields spectral index
$\beta_{\rm synch} = -2.5 \pm 0.1$
for emission traced by the synchrotron template,
with
free-free normalization
$\langle T_{\rm ff} / T_{\rm total} \rangle = 0.10 \pm 0.01$
~evaluated for latitudes $|b| < 10\arcdeg$
~in the lowest ARCADE 2 channel at 3.15 GHz.

The template model only specifies Galactic emission
up to an additive constant.
We fully specify the Galactic model
by computing the temperature of the template model
toward three reference locations
(north and south polar caps 
plus the coldest patch in the Northern hemisphere),
and adding a constant to the template model
to match the Galactic temperature 
derived by other means.
One estimate of the Galactic temperature
models the Galaxy as a simple plane-parallel structure
and derives the polar cap temperatures
from the slope of the Galactic emission
binned by the cosecant of Galactic latitude.
A second, independent method
computes the correlation between 
radio emission and 
C{\sc ii} line emission,
then estimates the radio brightness at each reference position
by multiplying the C{\sc ii} intensity at that position
by the measured radio/C{\sc ii} correlation slope.
Both methods yield similar estimates
for the total Galactic temperature towards each reference location.
We extend the analysis to full-sky surveys at lower frequencies
and find that Galactic polar cap temperature
from both methods
is consistent with a single power law
over the frequency range 22 MHz to 10 GHz,
with spectral index
$\beta = -2.55 \pm  0.03$
and normalization
$0.498 \pm 0.028$ K
at reference frequency $\nu_0$ = 1 GHz.

ARCADE2  produces well-calibrated maps of Galactic emission.
We use these maps to test for contributions
from spinning dust near the Galactic plane.
Previous tests for spinning dust at frequencies below 20 GHz
have used only the spatial correlation with template maps
tracing thermal dust emission.
The frequency dependence of the dust-correlated component
is not a stringent test for spinning dust,
since a weaker correlation at lower frequencies
is expected both for models with significant spinning dust contributions
as well as models with flat-spectrum synchrotron but no spinning dust.
We use a simple model of the total Galactic emission
to predict the mean intensity of the Galactic plane 
in the region observed by ARCADE 2.
The model normalizes the spinning dust contribution
at 22 GHz,
and computes the expected total contribution
from free-free, synchrotron, and spinning dust
as a function of frequency and spinning dust normalization.
The ARCADE 2 data consistently show less emission in the Galactic plane
than a model with no spinning dust,
and are consistent
with spinning dust contributing $0.4 \pm 0.1$ of the
Galactic plane emission at 22 GHz.

\acknowledgements

We thank the staff of the Columbia Scientific Balloon Facility
for their capable support throughout the integration, launch, flight, 
and recovery of the ARCADE 2 mission.
We thank the students whose work helped make ARCADE 2 possible:
Adam Bushmaker, 
Jane Cornett,
Sarah Fixsen,
Luke Lowe,
and
Alexander Rischard.
We thank P. Reich for providing the 45 MHz and 1420 MHz surveys
in electronic format.
We acknowledge use of the HEALPix software package.
This research is based upon work supported by
the National Aeronautics and Space Administration
through the Science Mission Directorate
under the Astronomy and Physics Research and Analysis 
suborbital program.
The research described in this paper
was performed in part at the Jet Propulsion Laboratory,
California Institute of Technology,
under a contract with the National Aeronautics and Space Administration.
T.V. acknowledges support from CNPq grants 466184/00-0, 305219-2004-9
and 303637/2007-2-FA, and the technical support from Luiz Reitano.
C.A.W. acknowledges support from CNPq grant 307433/2004-8-FA.


\end{document}